\documentclass{ws-p8-50x6-00}

\def\be{\begin{eqnarray}}
\def\en{\end{eqnarray}}
\def\non{\nonumber}
\def\la{\langle}
\def\ra{\rangle}
\def\nc{N_c^{\rm eff}}

\def\ov{\overline}

\def\pr{{\sl Phys. Rev.}~}
\def\prl{{\sl Phys. Rev. Lett.}~}
\def\pl{{\sl Phys. Lett.}~}
\def\np{{\sl Nucl. Phys.}~}
\def\zp{{\sl Z. Phys.}~}
\def\lsim{ {\ \lower-1.2pt\vbox{\hbox{\rlap{$<$}\lower5pt\vbox{\hbox{$\sim$}
}}}\ } }
\def\gsim{ {\ \lower-1.2pt\vbox{\hbox{\rlap{$>$}\lower5pt\vbox{\hbox{$\sim$}
}}}\ } }

\begin{document}

\title{Hadronic Two-Body Charmless $B$ Decays}
\author{Hai-Yang Cheng}

\address{Institute of Physics, Academia Sinica,
Taipei, Taiwan 115, Republic of China}


\maketitle

\abstracts{Implications of recent CLEO measurements of hadronic
charmless $B$ decays are discussed.}

\section{Effective Wilson Coefficients}
In the absence of first-principles calculations for hadronic
matrix elements, it is customary to evaluate the matrix elements
under the factorization hypothesis so that $\la O(\mu)\ra$ is
factorized into the product of two matrix elements of single
currents, governed by decay constants and form factors. However,
the naive factorized amplitude is not renormalization scale- and
$\gamma_5$ scheme- independent as the scale and scheme dependence
of Wilson coefficients are not compensated by that of the
factorized hadronic matrix elements. In principle, the scale and
scheme problems with naive factorization will not occur in the
full amplitude since $\la O(\mu)\ra$ involves vertex-type and
penguin-type corrections to the hadronic matrix elements of the
4-quark operator renormalized at the scale $\mu$. Formally, one
can write
\be
\la O(\mu)\ra=g(\mu,\mu_f)\la O(\mu_f)\ra,
\en
where $\mu_f$ is a factorization scale, and $g(\mu,\mu_f)$ is an
evolution factor running from the scale $\mu$ to $\mu_f$ which is
calculable because the infrared structure of the amplitude, if
any, is absorbed into $\la O(\mu_f)\ra$. Writing
\be
c^{\rm eff}(\mu_f)= c(\mu)g(\mu,\mu_f),
\en
the effective Wilson coefficients are formally scheme and
$\mu$-scale independent. The factorization approximation is then
applied to $\la O(\mu_f)\ra$ afterwards.

In principle, one can work with any quark configuration, on-shell
or off-shell, to compute the decay amplitude. If the off-shell
quark momentum is chosen as the infrared cutoff, $g(\mu,\mu_f)$
will depend on the gauge of the gluon field. This is all right as
the gauge dependence belongs to the infrared structure of the wave
function so that the physical amplitude is gauge independent
\cite{CLY}. However, if factorization is applied to $\la
O(\mu_f)\ra$, the information of gauge dependence characterized by
the wave function will be lost. Consequently, $c^{\rm eff}$ will
depend on the choice of gauge, a difficulty pointed out by Buras
and Silvestrini \cite{Buras98}. As pointed out in \cite{CLY},
within the factorization framework, one must work in the on-shell
quark scheme to obtain gauge invariant and infrared finite
$c_i^{\rm eff}$ and infrared poles, if any, are absorbed into
universal bound-state wave functions. It should be stressed that
the constant matrix $\hat r_V$ arising from vertex-like
corrections is {\it not} arbitrary due to the infrared finiteness
of vertex-like diagrams: The infrared divergences in individual
vertex-type diagrams cancel in their sum. The gauge-invariant
$\hat r_V$ matrices in naive dimension regularization  and 't
Hooft-Veltman renormalization schemes are first given in
\cite{CCTY} and \cite{CY10} (see also \cite{Beneke}).

\section{Nonfactorized Effects}
It is known that the effective Wilson coefficients appear in the
factorizable decay amplitudes in the combinations $a_{2i}=
{c}_{2i}^{\rm eff}+{1\over N_c}{c}_{2i-1}^{\rm eff}$ and
$a_{2i-1}= {c}_{2i-1}^{\rm eff}+{1\over N_c}{c}^{\rm eff}_{2i}$
$(i=1,\cdots,5)$. Phenomenologically, the number of colors $N_c$
is often treated as a free parameter to model the nonfactorizable
contribution to hadronic matrix elements and its value can be
extracted from the data of two-body nonleptonic decays.
Nonfactorizable effects in the decay amplitudes of $B\to PP,~VP$
can be absorbed into the parameters $a_i^{\rm eff}$. This amounts
to replacing $N_c$ in $a^{\rm eff}_i$ by $(N_c^{\rm eff})_i$.
Explicitly,
\be
a_{2i}^{\rm eff}={c}_{2i}^{\rm eff}+{1\over (N_c^{\rm
eff})_{2i}}{c}_{2i-1}^{ \rm eff}, \qquad \quad a_{2i-1}^{\rm eff}=
{c}_{2i-1}^{\rm eff}+{1\over (N_c^{\rm eff})_{2i-1}}{c}^{\rm
eff}_{2i},
\en
where $i=1,\cdots,5$ and
\be \label{nceff} (1/N_c^{\rm
eff})_i\equiv (1/N_c)+\chi_i\,,
\en
with $\chi_i$ being the nonfactorizable terms, which receive
contributions from nonfactorized vertex-type, penguin-type and
spectator corrections. In general, $\chi_i$ and $(\nc)_i$ are
complex.

Naive factorization does not work in the presence of nonfactorized
contributions. Nevertheless, if $\chi_i$ are universal (i.e.
channel by channel independent), then we still have generalized
factorization, which is likely to be justified for hadronic
charmless $B$ decays due to their large energy release. Since the
Fierz transformation of $(V-A)(V+A)$ operators is quite different
from that of $(V-A)(V-A)$ operators, we shall assume that
$\chi_{LR}\neq \chi_{LL}$, where $\chi_{LL}\equiv
\chi_{1,2,3,4,9,10}$ and $\chi_{LR}\equiv \chi_{5,6,7,8}$, or
equivalently,
 $N_c^{\rm eff}(LR)\neq N_c^{\rm eff}(LL)$ with
$N_c^{\rm eff}(LL)\equiv \left(N_c^{\rm
eff}\right)_{1,2,3,4,9,10}$ and $N_c^{\rm eff}(LR)\equiv
\left(N_c^{\rm eff}\right)_{5,6,7,8}$. As shown in \cite{CCTY},
the data analysis and the theoretical study of nonleptonic rare
$B$ decays all indicate that $\nc(LR)>3>\nc(LL)$. In principle,
$N_c^{\rm eff}$ can vary from channel to channel, as in the case
of charm decay. However, in the energetic two-body $B$ decays,
$\nc$ is expected to be process insensitive as supported by the
data \cite{CCTY}.

The observation $\nc(LL)<3<\nc(LR)$ is theoretically justified by
a recent perturbative QCD calculation of charmless $B$ decays in
the heavy quark limit. As pointed out
 in \cite{Beneke}, in
the heavy quark limit, all nonfactorizable diagrams are dominated
by hard gluon exchange, while soft gluon effects are suppressed by
factors of $\Lambda_{\rm QCD}/m_b$. In other words, the
nonfactorized term is calculable as expansion in $\alpha_s$ in the
heavy quark limit. Following \cite{Beneke}, we find the
nonfactorized terms:
\be
\chi_{LR}=-\chi_{LL}=-{\alpha_s\over 4\pi}\,{C_F\over N_c}(f^{\rm
I}+f^{\rm II}),
\en
where the hard scattering function $f^{\rm I}$ corresponds to hard
gluon exchange between the two outgoing light mesons and $f^{\rm
II}$ describes the hard nonfactorized effect involving the
spectator quark of the $B$ meson. Two remarks are in order. (i)
Since $f^{\rm I}$ is complex due to final-state interactions via
hard gluon exchange \cite{Beneke}, so are $\chi_i$ and $\nc(LL)$
and $\nc(LR)$. Nevertheless, the complex phases of $\chi_i$ are in
general small. (ii) Because Re\,$\chi_{LL}>0$, it is obvious that
$\nc(LL)<3$ and $\nc(LR)>3$. Furthermore, $\nc(LL)\sim 2$ implies
$\nc(LR)\sim 6$. Therefore, the empirical observation
$\nc(LR)>3>\nc(LL)$ shown in \cite{CCTY} gets a firm justification
from the perturbative QCD calculation.

\section{Tree-Dominated Charmless $B$ Decays}
CLEO has observed several tree-dominated charmless $B$ decays
which proceed at the tree level through the $b$ quark decay $b\to
u\bar ud$ and at the loop level via the $b\to d$ penguin diagrams:
$B\to \pi^+\pi^-,~\rho^0\pi^\pm,~\omega\pi^\pm,~\rho^\pm\pi^\mp$.
The updated branching ratios have been reported at this Conference
\cite{smith}: \be \label{pipi} {\cal B}(B^0\to\pi^+\pi^-) &=&
(4.3^{+1.6}_{-1.4}\pm 0.6)\times 10^{-6}, \non \\{\cal B}(B^\pm\to
\rho^0\pi^\pm) &=& (10.4^{+3.3}_{-3.4}\pm2.1)\times 10^{-6},
\non\\ {\cal B}(B^\pm\to \omega\pi^\pm) &=& (11.3^{+3.3}_{-2.9}\pm
1.5 )\times 10^{-6}, \non\\ {\cal B}(B^0\to\rho^\pm\pi^\mp) &=&
(27.6^{+8.4}_{-7.4}\pm 4.2)\times 10^{-6}.
\en
These decays are sensitive to the form factors $F_0^{B\pi}$,
$A_0^{B\rho}$, $A_0^{B\omega}$ and to the value of $\nc(LL)$. We
consider two different form-factor models for heavy-to-light
transitions: the BSW model \cite{BSW} and the light-cone sum rule
(LCSR) \cite{Ball} and obtain $1.1\leq \nc(LL)\leq 2.6$ from
$\rho^0\pi^\pm$ and $\omega\pi^\pm$ modes. This is indeed what
expected since the effective number of colors, $\nc(LL)$, inferred
from the Cabibbo-allowed decays $B\to (D,D^*)(\pi,\rho)$ is in the
vicinity of 2 (see \cite{CY}) and since the energy released in the
energetic two-body charmless $B$ decays is in general slightly
larger than that in $B\to D\pi$ decays, it is thus anticipated
that \be \label{chi} |\chi({\rm two-body~rare~B~decay})|\lsim
|\chi(B\to D\pi)|,
\en
and hence $\nc(LL)\approx \nc(B\to D\pi)\sim 2$.

Note that the branching ratio of $\rho^0\pi^\pm$ is sensitive to
the change of the unitarity angle $\gamma$, while $\omega\pi^\pm$
is not. For example, we have ${\cal B}(B^\pm\to\rho^0\pi^\pm)\sim
{\cal B}(B^\pm\to\omega\pi^\pm)$ for $\gamma\sim 65^\circ$, and
${\cal B}(B^\pm\to\rho^0\pi^\pm)> {\cal B}(B^\pm\to\omega\pi^\pm)$
for $\gamma>90^\circ$. It appears that a unitarity angle $\gamma$
larger than $90^\circ$, which is preferred by the previous
measurement \cite{Bishai} ${\cal B}(B^\pm\to \rho^0\pi^\pm)=(15\pm
5\pm 4)\times 10^{-6}$, is no longer strongly favored by the new
data of $\rho^0\pi^\pm$.

\section{$B\to\pi\pi$ and $\pi K$ Decays}
The CLEO measurement of $B^0\to\pi^+\pi^-$ mode [see Eq.
(\ref{pipi})] puts a very stringent constraint on the form factor
$F_0^{B\pi}$. Neglecting final-state interactions and employing
$\gamma\equiv{\rm Arg}(V_{ub}^*)=65^\circ$ and
$|V_{ub}/V_{cb}|=0.09$, and the effective number of colors
$\nc(LL)=2$, we find $F_0^{B\pi}(0)=0.20\pm 0.04$, which is rather
small compared to the BSW value\cite{BSW} $F_0^{B\pi}(0)=0.333$.
This relatively small form factor will lead to two difficulties.
First, the predicted $B\to K\pi$ branching ratios will be too
small compared to the data as their decay rates are governed by
the same form factor. Second, the predicted rate of $B\to K\eta'$
is also too small as the form factor $F_0^{BK}(0)$ cannot deviate
too much from $F_0^{B\pi}(0)$, otherwise the SU(3)-symmetry
relation $F_0^{BK}=F_0^{B\pi}$ will be badly broken.

There exist several possibilities that the $K\pi$ rates can be
enhanced: (i) The CKM matrix element $V_{ub}$ is small, say
$|V_{ub}/V_{cb}|\approx 0.06$, so that the form factor
$F_0^{B\pi}(0)$ is not suppressed \cite{Agashe}. However, this CKM
matrix element $|V_{ub}/V_{cb}|$ is smaller than the recent LEP
average \cite{LEP} $0.104^{+0.015}_{-0.018}$  and the CLEO result
\cite{Behrens2} $0.083^{+0.015}_{-0.016}$. (ii) A large nonzero
isospin $\pi\pi$ phase shift difference of order $70^\circ$ can
yield a substantial suppression of the $\pi^+\pi^-$
mode\cite{CCTY}. However, a large $\pi\pi$ isospin phase
difference seems to be very unlikely due to the large energy
released in charmless $B$ decays. Indeed, the Regge analysis of
\cite{Gerard} indicates $\delta_{\pi\pi}=11^\circ$. (iii) Smaller
quark masses, say $m_s(m_b)=65$ MeV, will make the $(S-P)(S+P)$
penguin terms contributing sizably to the $K\pi$ modes but less
significantly to $\pi^+\pi^-$ as the penguin effect on the latter
is suppressed by the quark mixing angles. However, a rather
smaller $m_s$ is not consistent with recent lattice calculations.
(iv) The unitarity angle $\gamma$ larger than $90^\circ$ will lead
to a suppression of $B\to\pi^+\pi^-$ \cite{gamma,CCTY}, which in
turn implies an enhancement of $F_0^{B\pi}$ and hence $K\pi$
rates. Therefore, the last possibility appears to be more
plausible.

We find that the CLEO $K\pi$ and $\pi\pi$ data can be accommodated
by $\gamma=105^\circ$,\footnote{Thus far, only the data of
$B\to\pi\pi,~K\pi$ imply the possibility that $\cos\gamma<0$. As
noted in Sec. III, $\gamma>90^\circ$ is not strongly favored by
the measurements of $B\to\rho^0\pi^\pm,~\omega\pi^\pm$.}
$F_0^{B\pi}(0)=0.28$, $F_0^{BK}(0)=0.36$, $\nc(LL)=2$. Note that
$\gamma=(114^{+25}_{-21})^\circ$ is obtained by Hou, Smith,
W\"urthwein \cite{HSW} under the assumption of naive
factorization. The calculated and experimental values of $K\pi$
decays are \be \label{Kpi}
 {\cal B}(\ov B^0\to K^-\pi^+) &=& 18.6\times 10^{-6}, \qquad
 (17.2^{+2.5}_{-2.4}\pm 1.2)\times
10^{-6}, \non \\ {\cal B}(B^-\to \ov K^0\pi^-) &=& 17.0\times
10^{-6}, \qquad (18.2^{+4.6}_{-4.0}\pm 1.6)\times 10^{-6}, \non\\
{\cal B}(B^-\to K^-\pi^0) & =& 12.6\times 10^{-6}, \qquad
(11.6^{+3.0+1.4}_{-2.7-1.3})\times 10^{-6}, \non\\  {\cal B}(\ov
B^0\to \ov K^0\pi^0) &=& 6.0\times 10^{-6}, \qquad~~
(14.6^{+5.9+2.4}_{-5.1-3.3})\times 10^{-6}.
\en
It is known that $K\pi$ modes are penguin dominated. As far as the
QCD penguin contributions are concerned, it will be expected that
${\cal B}(\ov B^0\to K^-\pi^+)\sim {\cal B}(B^-\to\ov K^0\pi^-)$
and ${\cal B}(B^-\to K^-\pi^0)\sim {\cal B}(\ov B^0\to\ov
K^0\pi^0)\sim {1\over 2}{\cal B}(B\to K\pi^\pm)$. However, as
pointed out in \cite{CCTY,gamma}, the electroweak penguin diagram,
which can be neglected in $\ov K^0\pi^-$ and $K^-\pi^+$, does play
an essential role in the modes $K\pi^0$. With a moderate
electroweak penguin contribution, the constructive (destructive)
interference between electroweak and QCD penguins in $K^-\pi^0$
and $\ov K^0 \pi^0$ renders the former greater than the latter;
that is, ${\cal B}(B^-\to K^-\pi^0)>{1\over 2}{\cal B}(\ov
B^0\to\ov K^0\pi^-)$ and ${\cal B}(\ov B^0\to\ov K^0\pi^0)<{1\over
2}{\cal B}(\ov B^0\to K^-\pi^+)$ are anticipated. We see from Eq.
(\ref{Kpi}) that, except for the decay $\ov K^0\pi^0$, the
agreement of the calculated branching ratios for $K\pi$ modes with
experiment is good. By contrast, the central value of ${\cal
B}(\ov B^0\to\ov K^0\pi^0)$ is much greater than the theoretical
expectation. Since its experimental error is large, one has to
await the experimental improvement to clarify the issue. The
predicted pattern $K^-\pi^+\gsim \ov K^0\pi^-\sim {3\over
2}K^-\pi^0\sim 3\,\ov K^0\pi^0$ is consistent with experiment for
the first three decays.

\section{$B\to K\phi$, $K\eta'$ and $K^*\eta$ Decays}
The decay amplitude of the penguin-dominated mode $B\to K\phi$ is
governed by $[a_3+a_4+a_5-{1\over 2}(a_7+a_9+a_{10})]$, where
$a_3$ and $a_5$ are sensitive to $\nc(LL)$ and $\nc(LR)$,
respectively. The current limit ${\cal B}(B^\pm\to\phi K^\pm)<
0.59\times 10^{-5}$ at 90\% C.L. \cite{Bishai} implies that  \be
\nc(LR)\geq \cases{5.0 & BSW, \cr 4.2 & LCSR, \cr} \en with
$\nc(LL)$ being fixed at the value of 2.  Hence, we can conclude
that $\nc(LR)>3>\nc(LL)$.

The improved measurements of the decays $B\to\eta' K$ by CLEO
yield \cite{smith}
\be  {\cal B}(B^\pm\to\eta' K^\pm) &=&
\left(80^{+10}_{-~9}\pm 7\right)\times 10^{-6}, \non \\
 {\cal B}(B^0\to\eta' K^0) &=& \left(89^{+18}_{-16}\pm 9
\right)\times 10^{-6}.
\en
This year CLEO has also reported the new measurement of $B\to
K^*\eta$ with the branching ratios \cite{Richichi} \be  {\cal
B}(B^\pm\to\eta K^{*\pm}) &=& \left(26.4^{+9.6}_{-8.2}\pm
3.3\right)\times 10^{-6}, \non \\
 {\cal B}(B^0\to\eta K^{*0}) &=& \left(13.8^{+5.5}_{-4.6}\pm 1.6
\right)\times 10^{-6}.
\en
Theoretically, the branching ratios of $K\eta'$ ($K^*\eta$) are
anticipated to be much greater than $K\pi$ ($K^*\eta'$) modes
owing to the presence of constructive interference between two
penguin amplitudes arising from non-strange and strange quarks of
the $\eta'$ or $\eta$. In general, the decay rates of $K\eta'$
increase slowly with $\nc(LR)$ if $\nc(LL)$ is treated to be the
same as $\nc(LR)$, but fast enough with $\nc(LR)$ if $\nc(LL)$ is
fixed at the value of 2. Evidently, the data much favor the latter
case. As stressed in \cite{CCTY}, the contribution from the
$\eta'$ charm content will make the theoretical prediction even
worse at the small values of $1/\nc$ if $\nc(LL)=\nc(LR)$\,! On
the contrary, if $\nc(LL)\approx 2$, the $c\bar c$ admixture in
the $\eta'$ will always lead to a constructive interference
irrespective of the value of $\nc(LR)$. The branching ratios of
$K^*\eta$ in general decrease with $\nc(LR)$ when
$\nc(LL)=\nc(LR)$ but increase with $\nc(LR)$ when $\nc(LL)=2$.
Again, the latter is preferred by experiment. Hence, the data of
both $K\eta'$ and $K^*\eta$ provide another strong support for a
small $\nc(LL)$ and for the relation $\nc(LR)>\nc(LL)$. In other
words, the nonfactorized effects due to $(V-A)(V-A)$ and
$(V-A)(V+A)$ operators should be treated differently.

Several new mechanisms have been proposed in the past few years to
explain the observed enormously large rate of $K\eta'$, for
example, the large charm content of the $\eta'$ \cite{Halperin} or
the two-gluon fusion mechanism via the anomaly coupling of the
$\eta'$ with two gluons \cite{Ahmady,Du}. These mechanisms will in
general predict a large rate for $K^*\eta'$ comparable to or even
greater than $K\eta'$ and a very small rate for $K^*\eta$ and
$K\eta$. The fact that the $K^*\eta$ modes are observed with
sizeable branching ratios indicates that it is the constructive
interference of two comparable penguin amplitudes rather than the
mechanism specific to the $\eta'$ that accounts for the bulk of
$B\to \eta' K$ and $\eta K^*$ branching ratios.

Finally, we would like to make a remark. As shown in \cite{CCTY},
the charged $\eta' K^-$ mode gets enhanced when $\cos\gamma$
becomes negative while the neutral $\eta' K^0$ mode remains
steady. Therefore, it is important to see if the disparity between
$\eta' K^\pm$ and $\eta' K^0$ is confirmed when experimental
errors are improved and refined in the future.


\end{document}